\documentclass[prb,a4paper,twocolumn,floatfix,showpacs,showkeys,amsmath,amssymb,nobibnotes,altaffilletter]{revtex4}

\usepackage{graphicx}
\usepackage{subfigure}
\usepackage{xspace}
\usepackage{pslatex}

\begin{document}

\preprint{}

\title{Influence of Charge Carrier Mobility on the Performance of Organic Solar Cells}

\author{Carsten Deibel$^1$, Alexander Wagenpfahl$^1$, Vladimir Dyakonov$^{1,2}$}
\affiliation{$^1$Experimental Physics VI, Julius-Maximilians-University of W{\"u}rzburg, D-97074 W{\"u}rzburg\\
$^2$Functional Materials for Energy Technology, Bavarian Centre for Applied Energy Research (ZAE Bayern), 
D-97074 W{\"u}rzburg}

%%PACS-Numbers
%71.23.An	Theories and models; localized states
% 72.20.Jv	Charge carriers: generation, recombination, lifetime, and trapping
% 72.80.Le	Polymers; organic compounds (including organic semiconductors)
%73.50.Pz	Photoconduction and photovoltaic effects
%73.63.Bd	Nanocrystalline materials


\begin{abstract}
The power conversion efficiency of organic solar cells based on donor--acceptor blends is governed by an interplay of polaron pair dissociation and bimolecular polaron recombination. Both processes are strongly dependent on the charge carrier mobility, the dissociation increasing with faster charge transport, with raised recombination losses at the same time. Using a macroscopic effective medium simulation, we calculate the optimum charge carrier mobility for the highest power conversion efficiency, for the first time accounting for injection barriers and a reduced Langevin-type recombination. An enhancement of the charge carrier mobility from $10^{-8}$m$^2$/Vs for state of the art polymer:fullerene solar cells to about $10^{-6}$m$^2$/Vs, which yields the maximum efficiency, corresponds to an improvement of only about 20\% for the given parameter set. \end{abstract}

\maketitle

Solution-processable organic photovoltaic cells have shown a promising performance increase in the recent years~\cite{scherf2008book}. A major focus now lies in finding a viable strategy for a further optimization of the power conversion efficiency, guided by a deeper understanding of the fundamental processes.  Device simulations are useful tools to assist in finding such routes, as they allow the extrapolation of possible but not yet implemented device concepts. We present macroscopic simulations of polymer--fullerene solar cells based on an effective medium approach. The influence of the charge carrier mobility on dissociation and transport processes, which are governing the power conversion efficiency, will be covered. Based on these results, we will discuss the most promising optimization routes for organic solar cells.

The macroscopic simulation program implemented by us solves the differential equation system of the Poisson, continuity and drift--diffusion equations by an iterative approach described by Gummel and Scharfetter~\cite{gummel1964,scharfetter1969}. Polaron pair dissociation~\cite{braun1984} as well as Langevin-type polaron recombination~\cite{langevin1908} are considered, as they are relevant for disordered organic semiconductors. In addition to previously published models \cite{mandoc2007}, we consider injection barriers by thermionic emission in order to reduce the unrealistically high carrier concentrations at the contacts, and also take the experimentally determined reduced Langevin recombination into account~\cite{juska2006,deibel2008mrs}. Therefore, we extend the Langevin recombination rate $R$ with a prefactor $\zeta$, 
\begin{equation}
	R = \zeta \gamma (n p - n_i^2 )
	\label{eqn:Langevin}
\end{equation}
where $n$ and $p$ are electron and hole concentration, respectively, $n_i$ is the (usually negligible) intrinsic carrier concentration, and $\gamma$ is the Langevin recombination parameter which is linearly proportional to the charge carrier mobility. The measure we use to quantify the influence of the bimolecular polaron recombination is defined as follows,
\begin{equation}
	\text{recombination yield} = 1 - \frac{U}{PG}
	\label{eqn:R}
\end{equation}
where $U=PG-(1-P)R$ is the net generation rate, $G$ is the exciton generation rate, and $P$ the polaron pair dissociation yield. We assume a small scale phase separation of donor and acceptor material, such that for each point within the active layer the exciton diffusion length is larger than the distance to the next acceptor site. Thus, we imply that all excitons are converted to polaron pairs, which is a reasonable assumption in the typical 1:1 donor--acceptor ratios~\cite{scherf2008book}. We note that the definition of a recombination yield is not unproblematic, as the net generation rate includes photogeneration, but recombination of photogenerated \emph{and} injected charge carriers. It is normalized to the rate of purely photogenerated polarons, $PG$. Surface recombination is implicitly assumed to be infinite, i.e., the electron and hole quasi Fermi levels meet at the electrodes.

\begin{table}[tb]
	\vspace*{-6mm}
	\caption{Parameters used for the macroscopic simulations.}\label{tab:param}
	\begin{tabular}{llrl}
		\hline
		\noalign{\vspace*{1mm}}
		Parameter\quad\quad  &Symbol\quad\quad  &Value\quad\quad  &Unit\\
		\hline
		temperature	 	& $T$ 			&$300$ &K \\
		effective band gap 	& $E_g$ 			&$1.35$ &eV \\
		dielectric constant	& $\epsilon$ 		& $3.4\cdot \epsilon_0$ & \\
		active layer thickness & $L$ 			& $115$ & nm \\
		effective density of states & $N_\text{eff}$ 	& $1\cdot 10^{27}$ & m$^{-3}$ \\
		generation rate		& $G$				& $4.15\cdot 10^{27}$ & m$^{2}$/s \\
		Langevin recombination \\
		\hspace{2ex} prefactor	& $\zeta$	& $1$ or $1/100$& \\
		polaron pair \\
		\hspace{2ex} separation length	& $a$	& $1.3$ & nm \\
		\hspace{2ex} recombination rate     & $k_f$	& $1\cdot 10^{4}$ & 1/s \\
		\hline
	\end{tabular}
	\vspace*{-1mm}
\end{table}

The parameters used in the simulations described here are summarized in Tab.~\ref{tab:param}, typical for annealed poly(3-hexyl thiophene):[6,6]-phenyl-C61 butyric acid methyl ester  (P3HT:PCBM, with ratio 1:1) devices. The reference mobility ratio between electrons and holes, $\mu_e$:$\mu_h$, is taken as 1, as transient photoconductivity measurements in this material blend showed mobilities of $10^{-8}$m$^2$/Vs for electrons and holes. These charge transport investigations will be published elsewhere.

\begin{figure}[tb]
	\center\includegraphics[height=70mm]{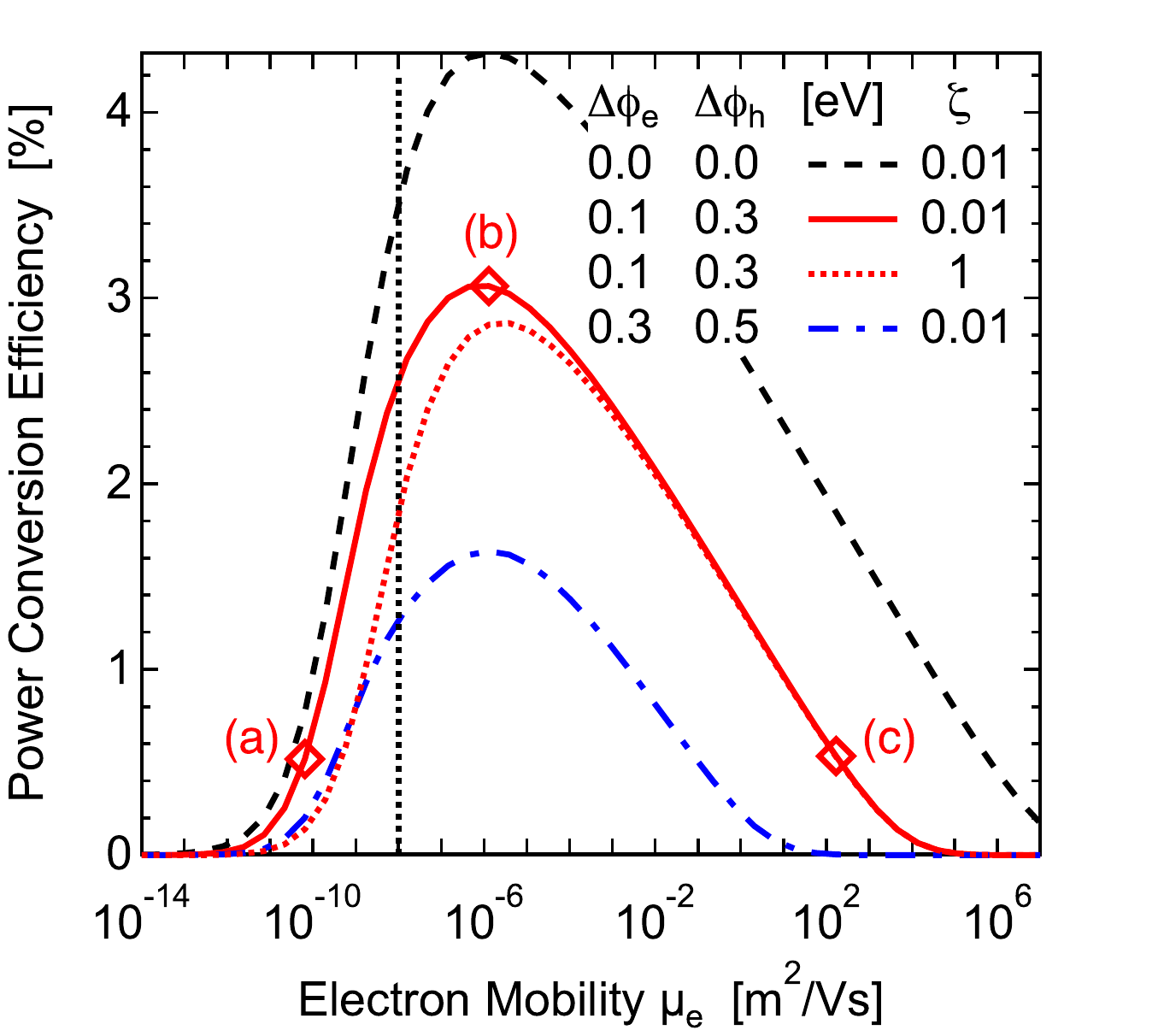}
	\caption{Power conversion efficiency of bulk heterojunction solar cell in dependence of charge carrier mobility, for different injection barriers at anode ($\Delta\Phi_h$) and cathode ($\Delta\Phi_e$). We assumed a reduced Langevin recombination by a factor $\zeta=1/100$, but included the still wide-spread $\zeta=1$ for comparison. The vertical dashed line indicates the experimentally determined electron and hole mobility of $10^{-8}$m$^2$/Vs for an annealed P3HT:PCBM (1:1) solar cell.}\label{fig: eta_vs_mu11}
\end{figure}

In Fig.~\ref{fig: eta_vs_mu11}, the power conversion efficiency of the bulk heterojunction solar cells versus the charge carrier mobility is shown. A pronounced maximum appears at a mobility of around $10^{-6}$m$^2$/Vs, its position being almost independent of the injection barrier height. Of course, a finite injection barrier leads to a somewhat lower efficiency. The point (b) represents the maximum power conversion efficiency, the two points (a)  and (c) denote 0.5\% efficiency.

\begin{figure}[tb]
	\center\includegraphics[height=70mm]{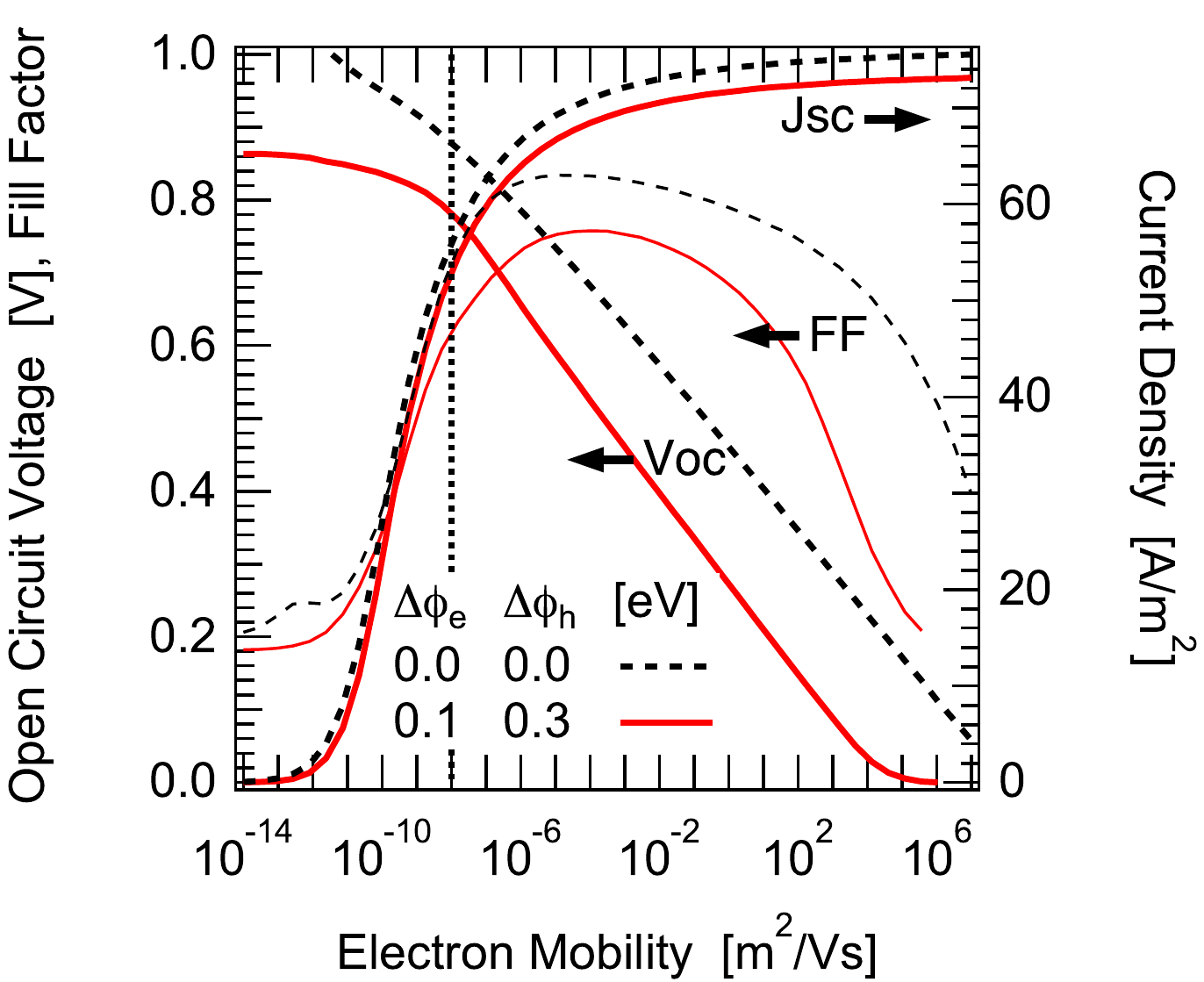}
	\caption{Open circuit voltage, fill factor and current density of the simulated solar cells in dependence of the charge carrier mobility,  for two sets of  injection barriers. The open circuit voltage grows with decreasing mobility.}\label{fig:SCparams_vs_mu_inj00_13}
\end{figure}

The corresponding solar cell parameters shown in Fig.~\ref{fig:SCparams_vs_mu_inj00_13} for two sets of  injection barriers already give some idea of the interplay responsible for the strong charge carrier mobility dependence of the performance. The open circuit voltage grows steeply with decreasing mobility. Considering injection barriers, it saturates at a value of $0.95$eV, given by the contact potential difference (band gap $E_g$ minus the electron and hole injection barriers, $ \Phi_e$ and $\Phi_h$). The photocurrent, in contrast, is maximum at the highest mobilities and decreases thereafter. The resulting mobility dependent power conversion efficiency is clearly a trade-off between open circuit voltage and photocurrent. 

\begin{figure}[tb]
	\center\includegraphics[height=70mm]{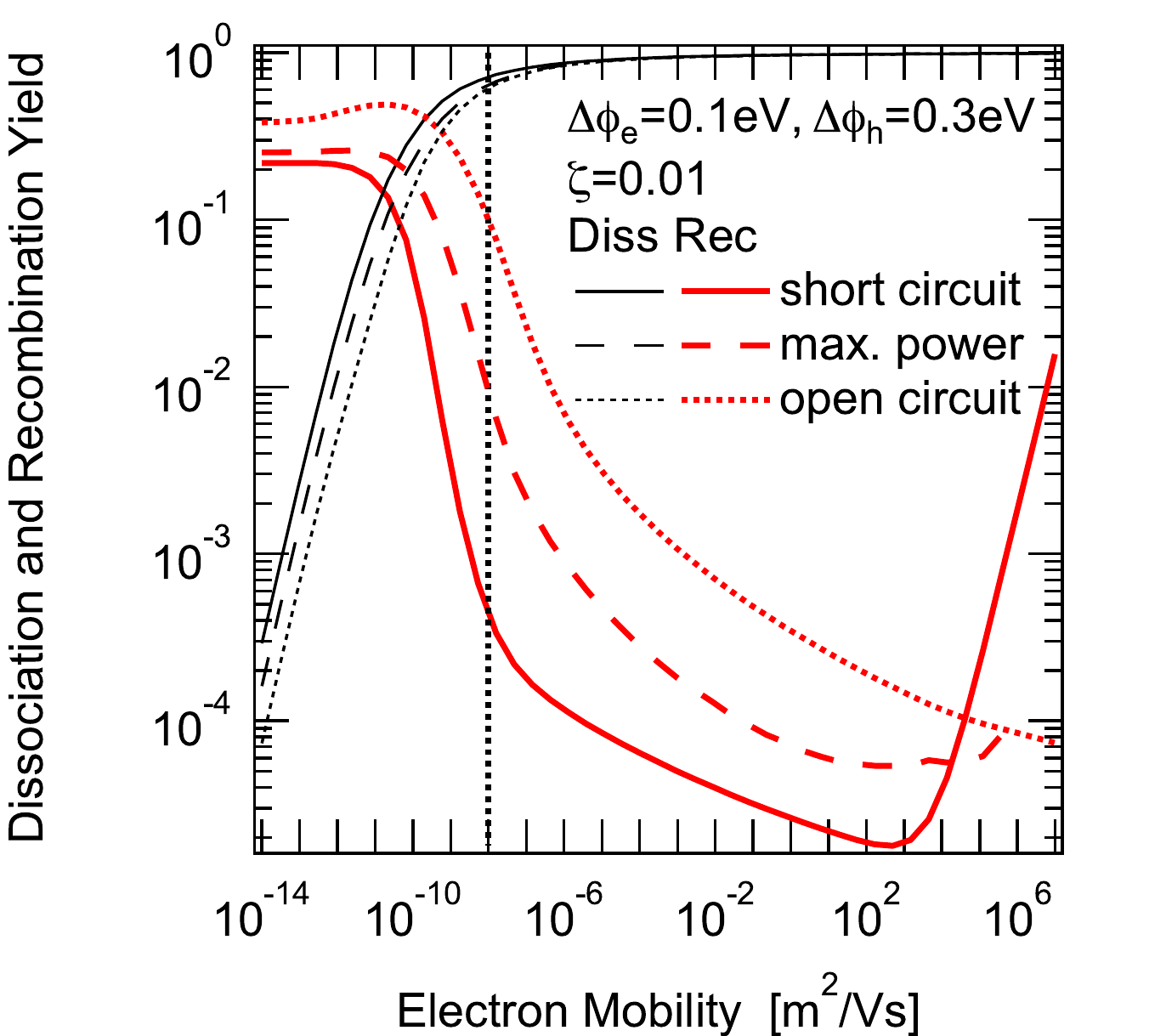}
	\caption{Polaron pair dissociation yield and recombination yield, for injection barriers of 0.1eV (cathode) and 0.3eV (anode), in dependence on the charge carrier mobility, under short circuit, maximum power point, and open circuit conditions. The vertical dashed line denotes a mobility of $10^{-8}$m$^2$/Vs. The dissociation yield grows with increasing mobility.}\label{fig: rec+diss_mu11_inj13}
\end{figure}

\begin{figure}[t]
	\center\includegraphics[height=70mm]{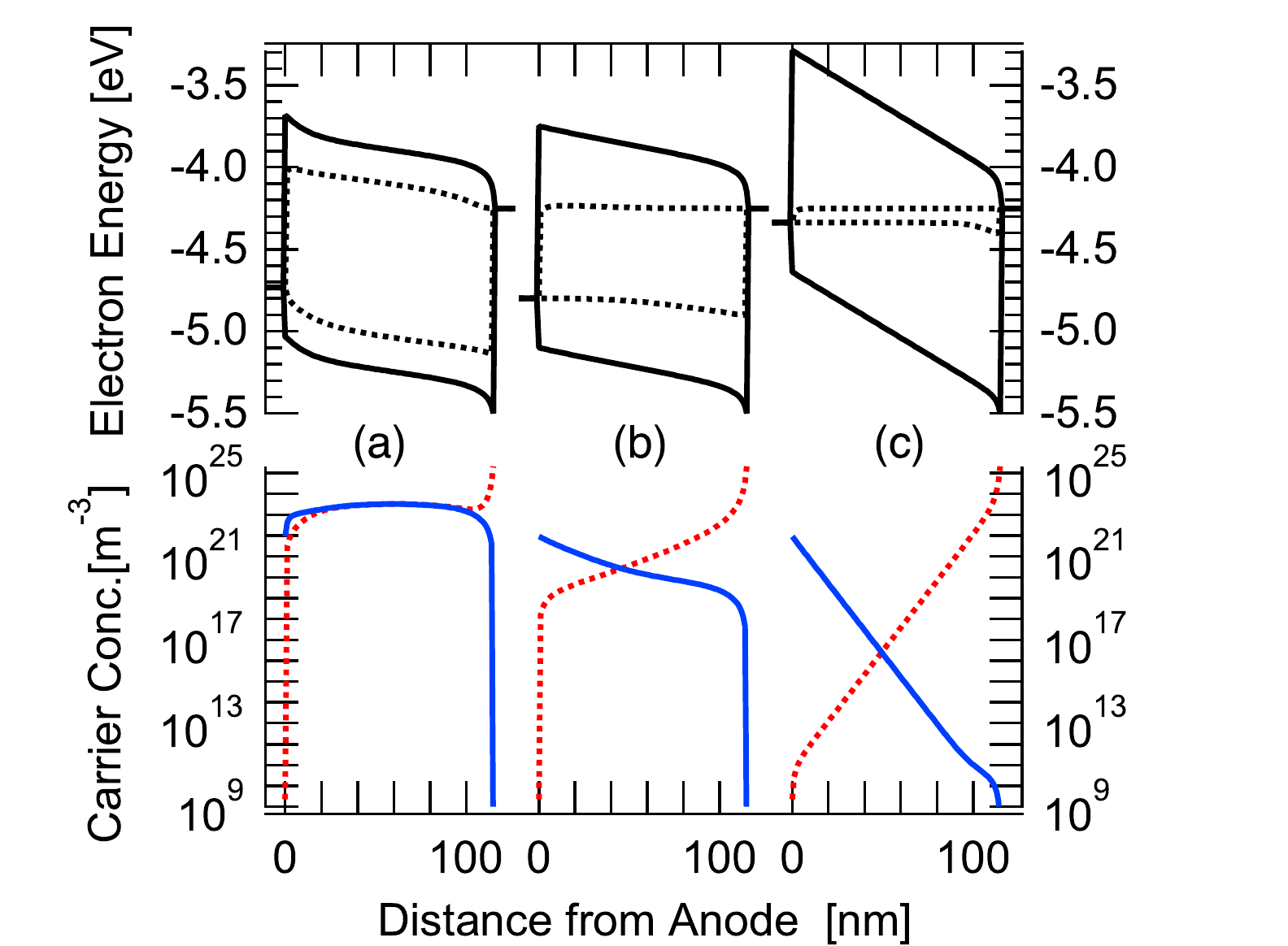}
	\caption{(Top) Band diagrams (solid) and electron and hole quasi-Fermi levels (dotted) as well as (bottom) electron (dotted) and hole (solid) concentrations at the maximum power point, for the three points indicated in Fig.~\ref{fig: eta_vs_mu11} by diamond symbols. The potential is normalized to the cathode work function of -4.3eV.}\label{fig:banddiagrams_mu11_inj13}
\end{figure}

The physical origin of these results becomes obvious when looking at the mobility dependent polaron pair dissociation yield $P$ and the nongeminate pair recombination yield $R$ (as defined in Eqn.~(\ref{eqn:R})), which can be seen in Fig.~\ref{fig: rec+diss_mu11_inj13}.  Already moderate mobilities lead to a very efficient separation of polaron pairs, as illustrated by the continuously growing dissociation yield. At low mobilities, the inefficient current extraction leads to a charge build-up. The latter is reflected by high carrier concentrations, as shown in Fig.~\ref{fig:banddiagrams_mu11_inj13} for the maximum power point, and yields an increased nongeminate recombination of free polarons (Eqn.~(\ref{eqn:Langevin})), rather than directly being dominated by the mobility. However,  the effect of the (reduced) Langevin recombination on the mobility dependent efficiency is not dominant, as shown in Fig.~\ref{fig: eta_vs_mu11}. Instead, the efficient charge extraction at high mobilities limits the open circuit voltage, as shown in Fig.~\ref{fig:banddiagrams_mu11_inj13}(c). This effect cannot be compensated by band bending at the organic--organic interface, as it is minimum in an effective medium such as a bulk heterojunction solar cell. In previously published analytical models~\cite{koster2005,cheyns2008}, a lower carrier concentration decreasing the open circuit voltage is well-established. However, we note that the steep open circuit voltage reduction is weakened if a finite surface recombination rate is considered.

We presented macroscopic simulations of the charge carrier mobility dependent performance of organic bulk heterojunction solar cells. The inclusion of the physically relevant injection barriers as well as an experimentally verified reduced Langevin recombination both lead to a lowered nongeminate recombination, which therefore plays only a small role for the device performance. The maximum attainable efficiency is a trade off between the polaron pair dissociation, which --- enhanced by a high carrier mobility --- results in high short circuit currents, and the efficient charge extraction at high carrier mobilities, which lead to low carrier concentrations, thus limiting the open circuit voltage. An enhancement of the charge carrier mobility from the state of the art of $10^{-8}$ to about $10^{-6}$m$^2$/Vs improves solar cell efficiency by only about 20\% for the given parameter set. For a further progress, novel device concepts --- improving the absorption by the use of strongly absorbing acceptor materials or tandem solar cells, and enhancing the polaron pair dissociation yield --- will have to be pursued.

\acknowledgments
C.D. thanks Uwe Rau for interesting discussions. V.D.'s work at the ZAE Bayern is financed by the Bavarian Ministry of Economic Affairs, Infrastructure, Transport and Technology.

\bibliographystyle{apsrev}

\begin{thebibliography}{[10]}
\bibitem{scherf2008book} U.~Scherf, C.~Brabec, V.~Dyakonov (Eds.), Organic Photovoltaics. Materials, Device Physics, and Manufacturing Technologies, Wiley VCH (2008).
\bibitem{gummel1964} H.K.~Gummel, IEEE Trans. Electron Dev. \textbf{ED11}, 455 (1964).
\bibitem{scharfetter1969} D.L.~Scharfetter, H.K.~Gummel, IEEE Trans. Electron Dev. \textbf{ED16}, 64 (1969).
\bibitem{braun1984} C.L.~Braun, J. Chem. Phys. \textbf{80} 4157 (1984).
\bibitem{mandoc2007} M. M. Mandoc, L. J. A. Koster, P. W. M. Blom, Appl. Phys. Lett. \textbf{90}, 133504 (2007).
\bibitem{langevin1908} P. Langevin, C. R. Acad. Sci (Paris) \textbf{146}, 530 (1908), translated by D.S.~Lemons and A.~Gythiel, Am. J. Phys. \textbf{65}, 1079 (1997).
\bibitem{juska2006} G. Juska, K. Arlauskas, J. Stuchlik, R. {\"O}sterbacka, J. Non-Cryst. Sol. \textbf{352}, 1167 (2006).
\bibitem{deibel2008mrs}  C.~Deibel, A.~Baumann, J.~Lorrmann, V.~Dyakonov, Mater. Res. Soc. Symp. Proc. \textbf{1031E}, 1031-H09-21 (2008).
\bibitem{koster2005}  L.~J.~A.~Koster, V.~D.~Mihailetchi, R.~Ramaker, P.~W.~M.~Blom, Appl. Phys. Lett.  \textbf{85}, 123509 (2005).
\bibitem{cheyns2008} D. Cheyns, J. Poortmans, P. Heremans, C. Deibel, S. Verlaak, B. P. Rand, and J. Genoe, Phys. Rev. B \textbf{77}, 165332 (2008).
\end{thebibliography}

\end{document}